\documentstyle[aps,preprint,tighten]{revtex}
\begin{document}

\title{Beyond the Narrow Resonance Approximation:  Decay
Constant and Width of the First Pion-Excitation State}

\author{Victor Elias and Amir H. Fariborz \\
Department of Applied Mathematics, University of Western Ontario \\
London, Ontario N6A 5B7 Canada}
\author{Mark A. Samuel \thanks{Permanent address: Department of
Physics, Oklahoma State University, Stillwater, Oklahoma 74078,
USA}\\
Department of Physics \\
McGill University \\
Montr\'eal, Quebec  H3A 2T8  Canada}
\author{Fang Shi and T. G. Steele\\
Department of Physics and Engineering Physics \\
University of Saskatchewan \\
Saskatoon, Saskatchewan S7N 5C6  Canada}

\date{June 23, 1997}

\maketitle

\begin{abstract}
      We consider the first pion excitation as a sub-continuum resonance
in the pseudoscalar channel, and we obtain parameters characterizing this
resonance through a global fit of the Borel-parameter dependence of the
field-theoretical pseudoscalar Laplace sum rule to its hadronic (pion +
pion-excitation + QCD-continuum) content.  Our analysis incorporates
finite-width deviations from the narrow resonance approximation,
instanton effects, and higher-loop perturbative contributions to the
pseudoscalar correlator.  We obtain the following values (uncertainties
reflect 90\% confidence levels):  mass M$_\Pi$ = 1.15 $\pm$ 0.28 GeV, width
$\Gamma_\Pi$ = $<$ 0.48 GeV, decay constant r $\equiv$
[F$_\Pi$M$_\Pi^2$ / f$_\pi$m$_\pi^2$]$^2$ = 4.7 $\pm$ 2.8.
\end{abstract}
\newpage
       
   The mass, width, and decay constant of the first pion excitation are
of interest to effective theories of low energy QCD.  Theoretical
arguments have been advanced to motivate masses below 1 GeV for this
state,$^{1,2}$ despite the M$_\Pi$ = 1300 $\pm$ 100 MeV entry in the 
present edition
of the Particle Data Guide.$^3$  The possibility that this resonance may be
very broad is also not excluded by present data, which gives a width
$\Gamma_\Pi$
between 200 and 600 MeV.$^3$  A broad width is suggestive of a small
value for this resonance's decay constant F$_\Pi$, which is experimentally
undetermined.  Recent theoretical work, however, suggests that
F$_\Pi$ may
be very small, even less than 1 MeV, if M$_\Pi$ is as large as 1300
MeV.$^2$
      If the first pion excitation is as light as 1 GeV, one cannot
automatically assume that this resonance's contribution to the correlation
function of pseudoscalar currents is absorbed in the continuum of
hadronic states approximated by purely-perturbative QCD,$^4$ a continuum
whose onset is generally assumed to be substantially above s = 1
GeV$^2$. $^5$ 
Consequently, a QCD sum-rule approach$^6$ to the I = 1 pseudoscalar
resonances cannot necessarily be limited to the lowest-lying resonance
(the pion).

      The idea of applying QCD sum-rule methods to the properties of
the first pion excitation is an old one.  Hubschmid and Mallik$^7$ included
the first pion-excitation state within a finite-energy sum-rule analysis of
the pseudoscalar channel, assuming a mass M$_\Pi$ = 1.5 GeV based upon
a linear Regge trajectory with a universal slope.  Their results are
suggestive (up to large uncertainties) of a substantially larger value for
F$_\Pi$ than anticipated from phenomenological hadronic physics.  Their
results, however, are obtained within the context of a narrow resonance
approximation in which $\Gamma_\Pi$ is assumed to be zero, as their 
analysis was
directed more towards a determination of quark masses than of
parameters characterizing the first pion excitation.

      The work presented here extracts information about the first pion
excitation's mass, decay constant, and width from a global fit of the
Borel transform of the pseudoscalar-current correlator to its hadronic
content, assuming that {\it both the pion and the first pion
excitation} reside
below the continuum threshold.  Subsequent pion excitations in our
analysis are assumed either to be above the continuum threshold, or to
have decay constants sufficiently small to decouple them from sub-
continuum 
physics.  We
depart from prior analyses $^{6,7,8,9,10,11}$ of the
pseudoscalar channel sum rules by incorporating 1) direct instanton
contributions, 2) finite-width effects, and 3) higher-order perturbative
contributions to the pseudoscalar current correlation function:

{\it 1. Direct Instanton Contributions:}  In the QCD sum-rule approach, 
long
distance effects are characterized by local matrix elements of quark and
gluon operators averaged over the physical vacuum -- the QCD-vacuum
condensates.  However, such {\it local} condensates, corresponding to 
vacuum
fluctuations of infinite correlation length, are insufficient to account 
for
the full nonperturbative content of the QCD vacuum, as they do not take
into account vacuum fluctuations of non-local origin arising from
instantons.$^{12}$  Such finite-correlation-length fluctuations may 
provide the
underlying mechanism for chiral symmetry breaking,$^{13}$ and need to be
separately included in an analysis of either scalar or pseudoscalar current
correlation functions.$^{10}$  The direct single-instanton contribution to 
the
Borel transform of the pseudoscalar correlation function has been
calculated in the instanton liquid model:$^{10,12}$

\begin{equation}
R_1^{inst}(\tau)  \equiv  4 m_q^2 \Pi^{inst} (\tau)  =
 \frac{3m_q^2 \rho_c^2}{2 \pi^2 \tau^3} \; e^{-\rho_c^2 / 2 \tau}
\left[ K_0 (\rho_c^2 / 2 \tau) + K_1 (\rho_c^2 / 2 \tau) \right],
\end{equation}

\noindent with instanton size denoted by $\rho_c$, and the light quark 
mass in the SU(2)-
symmetry limit denoted by m$_q$. Note that instanton effects can be
``turned off'' by letting $\rho_c \rightarrow \infty$. In eq. (1), also 
note that 
$\Pi(\tau)$ is the Borel
transform to the I = 1 component of the pseudoscalar-current correlation
function:

\begin{equation}
\Pi(\tau) = \frac{1}{\pi} \int_0^\infty Im [\Pi^p (s)] e^{-s\tau}
ds
\end{equation}

\begin{equation}
\Pi^p(q^2) = i \int d^4 x e^{iq \cdot x} < \Omega | T J^p (x) J^p (0)
| \Omega > ,
\end{equation}

\begin{equation}
J^p (x) = \frac{1}{\sqrt 2} [ \overline{u} (x) i \gamma_5 u(x) -
\overline{d} (x) i \gamma_5 d(x)].
\end{equation}

{\it 2. Finite-Width Effects:} Conventional methodology for determining the
hadronic content of QCD sum-rules rests upon the narrow resonance
approximation, in which hadronic Breit-Wigner contributions to the
imaginary parts of appropriate correlators are proportional to
$\delta$-functions:$^6$

\begin{equation}
\lim_{\Gamma \rightarrow 0} \; Im \left[ \frac{-1}{s - M^2 + i M \Gamma}
\right] = \pi \delta (s - M^2).
\end{equation}

\noindent One can undo the narrow resonance approximation through explicit
utilization of the Breit Wigner shape.$^{14}$ Analysis of the longitudinal
component of the axial-vector current correlator [corresponding to
4m$_q^2/s$
times the pseudoscalar current correlator] yields the following hadronic
contribution:

\begin{eqnarray}
R_1^{had}(\tau) = f_\pi^2 m_\pi^4 \; e^{-m_\pi^2 \; \tau} &+&  
\frac{F_\Pi^2 \; M_\Pi^2}{\pi}
\; \int_0^{s_0} \; \frac{M_\Pi
\; \Gamma_\Pi}{(s - M_\Pi^2)^2 + M_\Pi^2 \; \Gamma_\Pi^2} \; s\;
e^{-s\tau} ds \nonumber \\
& + & \frac{4 m_q^2}{\pi} \int_{s_0}^\infty \; Im \left[\Pi^p (s)
\right]^{pert} \; e^{-s\tau} \; ds.
\end{eqnarray}

\noindent The parameter $s_0$ represents the continuum threshold above 
which
hadronic physics coincides with purely-perturbative QCD. In the present
work, we first express the imaginary part of the Breit-Wigner shape
occurring in (6) as a Riemann sum of unit-area rectangular pulses
$P_m(s,\Gamma')$:
   
\begin{equation}
P_m (s, \Gamma') \equiv \left[ \Theta (s - m^2 +
m\Gamma') - \Theta (s - m^2 - m\Gamma') \right] / 2 m
\Gamma' ,
\end{equation}

\begin{eqnarray}
\frac{M \Gamma}{(s - M^2)^2+M^2 \Gamma^2}
= \lim_{n \rightarrow \infty} \frac{2}{n} \sum^n_{j=1}
\sqrt{\frac{n}{j-f} - 1} \; \; P_M (s, \; \sqrt{\frac{n}{j-f} - 1} \; \;
\Gamma) ,
\; \; \; [0 \leq f < 1].
\end{eqnarray}

\noindent A semi-analytic approximation for the integral over the Breit-
Wigner
shape can be realized by noting (for $s_0 > M_\Pi^2 + M_\Pi \Gamma'$) that
\begin{mathletters}
\begin{eqnarray}
\int_0^{s_0} P_{M_\Pi} (s, \Gamma') s\; e^{-s\tau} \; ds = M_\Pi^2
\; e^{-M_\Pi^2 \tau} \Delta (M_\Pi, \Gamma', \tau), 
\label{9a} 
\end{eqnarray}

\begin{eqnarray}
\Delta (M, \Gamma, \tau) \equiv \frac{ \sinh (M \Gamma \tau)}{M
\Gamma \tau} \left[ 1 + \frac{1}{M^2 \tau} \right] - \frac{\cosh (M
\Gamma \tau)}{M^2 \tau} \; \;  \begin{array}{c}  {}  \\  \longrightarrow \\
\Gamma \rightarrow 0 \end{array} \; \; 1.
\label{9b}
\end{eqnarray}
\end{mathletters}
\noindent In the present analysis, we have utilized a 4-pulse 
approximation to the
right-hand side of (8), choosing f = 0.7 in order that the area of the four
pulses be equivalent to the total area (= $\pi$) under the Breit-Wigner
curve.
[In the $n \rightarrow \infty$ limit, such equivalence is true for any 
choice of f between
0 and 1].  We then obtain a simple expression for the finite-width
correction W to the narrow-resonance-approximation contribution: 

\begin{equation}
\frac{F_\Pi^2 M_\Pi^2}{\pi} \int_0^{s_0} \frac{M_\Pi \Gamma_\Pi}{(s -
M_\Pi^2)^2 + M_\Pi^2 \Gamma_\Pi^2} \; s \; e^{-s \tau} \; ds =
F_\Pi^2 M_\Pi^4 \; e^{-M_\Pi^2 \tau} \; W [M_\Pi, \Gamma_\Pi, \tau]
\end{equation}

\begin{eqnarray}
W [M, \Gamma, \tau] & = & 0.5589 \Delta (M, 3.5119 \; \Gamma, \tau)
+  0.2294 \Delta (M, 1.4412 \; \Gamma, \tau) \nonumber \\
& + & 0.1368 \Delta (M, 0.8597 \; \Gamma, \tau) + 0.0733 \Delta (M, 0.4606 
\; \Gamma, \tau).
\end{eqnarray}

\noindent Note that $W \rightarrow 1$ in the narrow-resonance limit
($\Gamma_\Pi \rightarrow 0$).  The four-
pulse approximation we utilize serves to mitigate numerical difficulties 
that arise
from the infinite Breit-Wigner tail, which extends unphysically into both
Euclidean (s $<$ 0) and post-continuum (s $>$ s$_0$) values of s.

{\it 3. Higher-Order Perturbative Contributions:}  Virtually all previous 
sum-
rule treatments of the pseudoscalar channel have made use only of the
one-loop expression [or incomplete 2-loop expressions] for the purely
perturbative QCD-contribution to the pseudoscalar current correlator.  
However,
the perturbative content of the pseudoscalar current correlation function 
has been
known for some time to three-loop order, with higher-loop contributions
accompanied by numerically large coefficients:$^{15}$

\begin{eqnarray}
Im [\Pi^p (s)]^{pert}  
& = &  \frac{3s}{8 \pi} \left\{ 1 + \frac{\alpha_s}{\pi} \left[
\frac{17}{3} - 2 \ln (s \tau) \right] \right. \nonumber \\
& + & \left. \left( \frac{\alpha_s}{\pi} \right)^2 \left[ 40.684 - 31.667
\ln (s \tau) - 1.417 \pi^2 + 4.251 \ln^2 (s \tau) \right] +  O
(\alpha_s^3) \right\}.
\end{eqnarray}

\noindent The renormalization point $\mu = \tau^{-1/2}$ is chosen to 
identify the Borel
parameter as the renormalization-group (RG) scale parameter. In the
work presented here, we have considered both 2- and 3-loop expressions
for $\Pi^p(s)$ in determining parameters characterizing the first pion
excitation.  Computational limitations prevented us from determining
90\% confidence-level bounds on our fitted parameters when using the 3-
loop expression--we could obtain numerically efficient Monte Carlo 
estimates of these
confidence levels only for the 2 loop case.  We were, however, able to
compare explicitly the $\chi^2$-minimizing fits obtained when either 2- or 
3-
loop expressions for $\Pi^p(s)$ are utilized. Rows 5 and 6 of Table I
demonstrate that the pion-resonance mass, decay constant,
and width are shifted very little in going from two to three loops,
despite potentially large 3-loop perturbative corrections. It should
be noted that the minimum $\chi^2$ increases by an order of magnitude
when the pion resonance is omitted (leaving only the ground state
pion).  This indicates that, unlike the case for other channels, the
pion resonance is too strong to be absorbed into the continuum.

      Our analysis of pion-resonance properties follows from a weighted
least squares fit of the $\tau$-dependence [0.4 GeV$^{-2} \leq \tau \leq
2.5$ GeV$^{-2}$] of the
hadronic contributions to R$_1(\tau)$ to the corresponding field 
theoretical
expression, as derived from the pseudoscalar current correlation function. 
 
The region of $\tau$ is chosen by placing a 20\% upper bound on the
theoretical uncertainty in $R_1(\tau)$, based upon a 30\% continuum
and a 50\% power-law-uncertainty. The
hadronic contribution, which explicitly contains the fitted pion-resonance
parameters M$_\Pi$, F$_\Pi$, $\Gamma_\Pi$ and the continuum threshold
s$_0$, is obtained via
substitution of eqs (10-12) into (6).  The field-theoretical contribution
$R_1^{ft}(\tau) [= 4m_q^2 \Pi(\tau)$ as defined by eq. (2)] contains 
direct instanton
contributions [eq. (1)], QCD-vacuum condensate
contributions,$^{6,8,16}$ and
purely-perturbative contributions that can be extracted by substitution of
(12) into (2). Since the quantity $R_1^{ft}(\tau)$ satisfies a 
renormalization group
equation,$^{17}$ coupling constants $\alpha_s$ and quark masses $m_q$ 
appearing
outside RG-invariant condensates can be replaced with scale-parameter
dependent RG-improved expressions referenced to the $\mu =
\tau^{-1/2}$ mass
scale to the appropriate loop level.  The two loop expressions are:

\begin{equation}
\alpha_s(\tau) = \frac{2 \pi}{9L} \left[ 1 - \frac{32 \;
\ln(L)}{81 \; L} \right] \; \; , \; L \equiv -\frac{1}{2} \ln (\tau
\Lambda^2) , 
\end{equation}
\begin{equation}
m(\tau) = \hat{m} L^{-4/9} \left[ 1 - \frac{0.1989 - 0.1756 \ln
(L)}{L} \right] \equiv  \hat{m} \zeta(\tau) .
\end{equation} 

\noindent The parameter $\hat m$  is the RG-invariant quark mass, which we 
include as
one of our fitted parameters by comparison of the explicit $\tau$-
dependence
of

\begin{equation}
R_1^{had}(\tau) / \hat{m}^2 =  a \left[ 1 + re^{-M_\pi^2 \tau} W \left[ 
M_\Pi, \Gamma_\Pi,
\tau \right] +  \frac{4 \zeta^2 (\tau)}{\pi} \int_0^{s_0} Im \left[
\Pi^p (s)
\right]^{pert} e^{-s \tau} \; ds \right],
\end{equation}

\begin{equation}
a = f_\pi^2 m_\pi^4 / \hat{m}^2, \; \; \; r = \left( F_\Pi M_\Pi^2 /
f_\pi m_\pi^2 \right)^2 .
\end{equation}

\noindent to the corresponding field-theoretical expression:

\begin{eqnarray}
 R_1^{ft}(\tau) / \hat{m}^2 & = & \zeta^2 (\tau) \left\{ \frac{3}{2 \pi^2 
\tau^2}  \left[ 
1 +  \frac{17 \alpha_s (\tau)}{3 \pi} + \frac{26.699 \alpha_s^2
(\tau)}{\pi^2} + \;  (1 - \gamma_E) \left[ -2 \alpha_s (\tau) / \pi - 
31.667
\alpha_s^2 (\tau) / \pi^2 \right] \right. \right. \nonumber \\
& + & \left. \left( \frac{\pi^2}{6} + \gamma_E^2 - 2 \gamma_E \right) 
\left(
\frac{4.251 \alpha_s^2 (\tau)}{\pi^2} \right) \right] -  4 < m \bar{q}q> + 
\frac{1}{2 \pi} <
\alpha_s G^2> + \frac{448 \pi
\tau}{27} < \alpha_s (\bar{q}q)^2 > \nonumber \\
& + & \left. \frac{3\rho_c^2}{2 \pi^2 \tau^3} e^{-\rho_c^2 / 2 \tau} 
\left[ K_0
\left( \rho_c^2 / 2 \tau \right) + K_1 \left( \rho_c^2 / 2 \tau
\right) \right] 
+ O(m_q) \right\}
\end{eqnarray}

\noindent SU(2) breaking effects are higher order in m$_q$, since their 
contributions
to R$_1$ are in terms proportional to $(m_u - m_d)^2(m_u +
m_d)^2$.$^8$  The
contributions to R$_1$ from the dimension-5 quark-antiquark-gluon
condensate and the dimension-6 triple-gluon condensate are proportional
to $m_q^3$,$^{16}$ and are therefore not included in (17).  Our fit is 
obtained using
a standard set of perturbative and nonperturbative RG-(quasi-)invariant
QCD parameters: $<m\bar{q}q> = -f_\pi^2 m_\pi^2/4$ [$f_\pi$ = 131 MeV],
$<\alpha_sG^2>$ = 0.045 GeV$^4$,
$\Lambda$ = 0.15 GeV,
$\rho_c$ = (600 MeV)$^{-1}$. 
Input assumptions concerning the deviation of the dimension-6 fermion
condensate $<\alpha_s (\bar{q}q)^2>$ from its standard$^6$ value
$1.8 \cdot 10^{-4}$ GeV$^6 \equiv S_6$ are specified in Table I,
and are discussed below.
Optimal
values for the parameters a, r, M$_\Pi$, $\Gamma_\Pi$, and $s_0$ 
correspond to the fit of the
$\tau$ dependence of (15) to (17) that minimizes a $\chi^2$ weighted for 
the previously described 50\%
uncertainty for power-law corrections and a 30\% uncertainty for
continuum corrections.  As mentioned earlier, uncertainties in the fitted
parameters are obtained only for the two-loop case [for which the integral
in (15) can be easily determined analytically] from a Monte-Carlo
simulation incorporating the power-law and continuum uncertainties
described above, as well as a 15\% variation in the value of
$\rho_c$, and (in rows 1-4 of Table I), a factor of two vacuum-saturation 
uncertainty in the value $S_6$ for $< \alpha_s (\bar{q}q)^2>$.

The value $S_6$ was originally obtained $^6$ through use of vacuum
saturation [$<(\bar{q}q)^2> = <\bar{q}q>^2$] in conjunction with the
value for $<m \bar{q}q>$ appropriate for a 5 MeV quark mass.  Either
a heavier quark mass and/or a direct violation of the
vacuum-saturation assumption contribute substantial uncertainty to
this quantity, which our Monte-Carlo simulation allows to vary over a
factor of 2.  To demonstrate the overall insensitivity of our fitting
procedure to the value of $<\alpha_s (\bar{q}q)^2>$, we have
constrained this condensate to equal exactly the standard value $S_6$
in rows 5 and 6 of Table I.  Comparison of rows 4 and 5
demonstrates that removal of the Monte-Carlo variation in the
magnitude of the dimension-6 condensate has virtually no effect on
our results.  The results in rows 5 and 6, however, are indicative of a 
substantial violation of
vacuum-saturation:  the standard parameter value $S_6$ is much larger
than that anticipated for $<\alpha_s (\bar{q}q)^2>$ via vacuum
saturation if the quark mass at $\mu = 1$ GeV is of order 8-9 MeV. 
Row 7 of the Table corresponds to a fit in which exact
vacuum-saturation {\it is imposed} on the value of $<\alpha_s
(\bar{q}q)^2>$, by including the implicit dependence of this quantity
on the fitted RG-invariant quark mass $\hat{m}$.  Thus, the
requirements $<(\bar{q}q)^2> = <\bar{q}q>^2$ and $<\bar{q}q> =
-f_\pi^2 m_\pi^2 / 4 m_q$ are simultaneously upheld.  When included
in our fitting procedure, these input assumptions lead to a $\mu = 1$ GeV
quark mass much larger than 5 MeV, although the upper end of the
range obtained for $M_\Pi$ remains consistent with the 1200 MeV
empirical lower bound.$^3$

To conclude, 90\% confidence-level results for the first
pion-excitation are summarized in rows 4-7 of Table I.  These
results incorporate 1) the explicit direct-instanton contribution, 2) 
explicit finite-width effects, 3) the full two- and three-loop perturbative
contribution to the sum rule, and 4) the violation [or explicit
preservation (row 7)] of the vacuum saturation assumption for the
dimension-6 fermionic condensate.  We believe the results of row 4 to
be phenomenologically salient, as the fitted range for $M_\Pi$ is found to 
be fully
inclusive of the present 1300 $\pm$ 100 MeV empirical range.$^3$ 
The final three rows suggest a somewhat lighter pion-excitation than 
that indicated in
the Particle Data Guide, though a mutually consistent value of 1200
MeV is not ruled out. 

Although the results of rows 4-7 vary according to the input
assumptions employed, a number of common features are evident.
The decay constant F$_\Pi$ [using PDG conventions
consistent with f$_\pi$= 131 MeV] is seen to be at least 2.8 MeV,
consistent with the larger values of $F_\Pi$ already anticipated by
Hubschmid and Mallik.$^7$  Although little sensitivity is evident to
the fitted value of the width (the upper bounds quoted are 90\%
confidence levels), values larger than 480 MeV for $\Gamma_\Pi$ appear to 
be ruled out. The fits we obtain are found (surprisingly) to uphold
the narrow resonance aproximation ($\Gamma = 0$).  The fitted
parameter {\it a} allows a determination  of the RG-invariant quark
mass $\hat{m}$, and a subsequent determination via (14) of the SU(2)-
symmetric quark
mass at a $\mu$ = 1 GeV momentum scale.  This value in row 4
is less than values anticipated in the absence of
instanton corrections (rows 1 and 3). 
A comparison of rows 3 and 4 also demonstrates that direct
instanton contributions reduce somewhat the value of M$_\Pi$, but
increase F$_\Pi$.  Finally, we note the relative insensitivity
in Table I to the value of the continuum threshold $s_0$.  
The analysis
clearly favours a threshold above $M_\Pi^2$ for the onset of equivalence 
between
perturbative and hadronic QCD.  The viability of larger values for
$s_0$
suggests that smaller values for the pion-excitation decay
constant $F_\Pi$ could
be most easily accommodated by incorporating additional pion
resonances into sub-continuum physics. 

\bigskip

\begin{section}*{Acknowledgements}
V.E. is grateful for discussions with A. L. Kataev, V. A. Miransky, and
A. I. Vainshtein.  M. A. S. would like to thank the Department of
Applied Mathematics of the University of Western Ontario for its kind
hospitality. This research has been supported by grants from the
Natural Sciences and Engineering Research Council of Canada.
\end{section}

\bigskip

\begin{section}*{References}
\begin{enumerate}
\item A. A. Andrianov and V. A. Andrianov, Nucl. Phys. B (Proc.
Supp't) {\bf 39 B, C}, 257 (1995).

\item M. K. Volkov and C. Weiss, 1996 preprint hep-ph/9608347
(unpublished).

\item R. M. Barnett et. al. [Particle Data Group], Phys. Rev. {\bf
D54}, 1 (1996).

\item J. J. Sakurai, Phys. Lett. {\bf 46B}, 207 (1973).

\item L. J. Reinders, H. Rubinstein, and S. Yazaki, Phys. Rep. {\bf
127}, 1, (1985).

\item M. A. Shifman, A. I. Vainshtein, and V. I. Zakharov, Nucl.
Phys. {\bf B147}, 385 and 448 (1979).

\item W. Hubschmid and S. Mallik, Nucl. Phys. {\bf B193}, 368 (1981).

\item C. Becchi, S. Narison, E. de Rafael, and F. J. Yndurain, Zeit.
Phys. {\bf C8}, 335 (1981).

\item V. A. Novikov, M. A. Shifman, A. I. Vainshtein, and V. I.
Zakharov, Nucl. Phys. {\bf B191}, 301 (1981).

\item E. V. Shuryak, Nucl. Phys. {\bf B214}, 237 (1983).

\item J. Gasser and H. Leutwyler, Phys. Rep. {\bf 87}, 77 (1982).

\item A. E. Dorokhov, S. V. Esaibegyan, N. I. Kochelev, and N. G.
Stefanis, 1996 preprint hep-th/9601086 (unpublished).

\item R. D. Carlitz and D. B. Creamer, Ann. Phys. {\bf 118}, 429
(1979);  D. I. Diakonov and V. Yu. Petrov, Zh. Eksp. Teor. Fiz. {\bf
89}, 361 and 751 (1985).

\item C. A. Dom\'inguez, Zeit. Phys. {\bf C26}, 269 (1984).

\item S. G. Gorishny, A. L. Kataev, and S. A. Larin, Phys. Lett. {\bf
135B}, 457 (1984).

\item E. Bag\'an, J. I. LaTorre, and P. Pascual, Z. Phys. {\bf C32},
43 (1986).

\item S. Narison and E. de Rafael, Phys. Lett. {\bf 103B}, 57 (1981).
\end{enumerate}
\end{section}

\newpage

\begin{center}
Table I
\end{center}

\baselineskip=16pt
\noindent
\begin{tabular}{|l|l|l|l|l|l|l|l|l|l|l|l|r|} \hline
\multicolumn{4}{|c|} {INPUTS} &
\multicolumn{7}{c|} {OUTPUTS} \\ \hline
$\rho_c(MeV^{-1})$ & $\Gamma$ & $l$ & $<\alpha_s (\bar{q}q)>^2$ & 
$M_\Pi$(GeV) & $r$
& $F_\Pi$ (MeV) & $a$
(GeV$^4$) & $m_q$[$\mu$=1GeV] & $\Gamma$ (GeV) & $s_0$ (GeV$^2$) \\
\hline \hline
$\infty$ & 0 & 2 & $0-2S_6$ & 1.34 $\pm$ 0.16 & 8.9 $\pm$ 3.2 & 4.3 $\pm$
1.3 &
0.049 $\pm$ 0.016 & 9.1 $\pm$ 1.5 MeV & -- & 3.4 $\pm$ 1.4 \\
\hline
1/600 & 0 & 2 & $0-2S_6$ & 1.0 $\pm$ 0.25 & 5.4 $\pm$ 3.8 & 6.0 $\pm$
3.7 &
0.063 $\pm$ 0.035 & 8.0 $\pm$ 2.2 MeV & -- & 3.2 $\pm$ 1.5 \\ 
\hline
$\infty$ & -- & 2 & $0-2S_6$ & 1.31 $\pm$ 0.25 & 5.7 $\pm$ 2.8 &
3.6 $\pm$ 1.6 &
0.048 $\pm$ 0.014 & 9.2 $\pm$ 1.3 MeV & $<$0.50 & 2.7 $\pm$
1.2 \\ \hline
1/600 & -- & 2 & $0-2S_6$ & 1.15 $\pm$ 0.28 & 4.7 $\pm$ 2.8 & 4.2 $\pm$
2.4 &
0.069 $\pm$ 0.037 & 7.7 $\pm$ 2.2 MeV & $<$0.48 & 3.7 $\pm$ 1.2
\\ \hline 
1/600 & -- & 2 & $S_6$ & 1.07 $\pm$ 0.17 & 4.6 $\pm$ 1.8 & 4.8
$\pm$ 1.8 & 0.052 $\pm$ 0.014 & 8.9 $\pm$ 1.2 MeV & $<$0.48 &
3.6 $\pm$ 1.4 \\ \hline
1/600 & -- & 3 & $S_6$ &1.05 & 4.36 & 4.86 & 0.0622 & 8.10 MeV & 0 &
2.94 \\ \hline 
1/600 & -- & 2 & S$_6$(5.3/$\hat{m})^2$ & 0.95 $\pm$ 0.23 & 6.3 $\pm$
3.8 & 7.2 $\pm$4.1 & 0.036 $\pm$ 0.019 & 10.6 $\pm$ 2.8 MeV & $<$
0.30 &
3.0 $\pm$ 1.0 \\ \hline \hline
\end{tabular}

\bigskip

\noindent Fit of $\Pi$(1300) pion-resonance parameters to 90\%
confidence levels under various input assumptions:

\bigskip

\noindent $\rho_c = \infty$:  No direct-instanton contribution.

\noindent $\Gamma = 0$:  Narrow resonance approximation.

\noindent $l = 2$:  Full two-loop perturbative contribution.

\noindent $l = 3$:  Full three-loop perturbative contribution to the
sum rule.  For this case, only the fit which minimizes $\chi^2$ is
quoted.
\indent Computational limitations precluded the determination of
90\% confidence levels.

\noindent $S_6$: $S_6 \equiv 1.8 \cdot 10^{-4}$ GeV$^6$ (see text),
corresponding to exact vacuum saturation when $\hat{m} \cong 5$ MeV. 
Row 7 corresponds to
\indent exact vacuum-saturation of $<\alpha_s(\bar{q}q)^2>$ for the {\it 
fitted
values} of $\hat{m}, \; m_q$.

\end{document}